# Contact Research Strategy for Emerging Molybdenum Disulfide and Other Two-Dimensional Field-effect Transistors

Yuchen Du, Lingming Yang, Han Liu, and Peide D. Ye [a]

*School of Electrical and Computer Engineering and Birck Nanotechnology Center,*

*Purdue University, West Lafayette, IN 47907, U.S.A.*

Layered two-dimensional (2D) semiconducting transition metal dichalcogenides (TMD) have been widely isolated, synthesized, and characterized recently. Numerous 2D materials are identified as the potential candidates as channel materials for future thin film technology due to their high mobility and the exhibiting bandgaps. While many TMD filed-effect transistors (FETs) have been widely demonstrated along with a significant progress to clearly understand the device physics, large contact resistance at metal/semiconductor interface still remain a challenge. From 2D device research point of view, how to minimize the Schottky barrier effects on contacts thus reduce the contact resistance of metals on 2D materials is very critical for the further development of the field. Here, we present a review of contact research on molybdenum disulfide and other TMD FETs from the fundamental understanding of metal-semiconductor interfaces on 2D materials. A clear contact research strategy on 2D semiconducting materials is developed for future high-performance 2D FETs with aggressively scaled dimensions.

[a] Author to whom correspondence should be addressed; electronic mail: yep@purdue.edu

As the forefather of the layered 2D materials, graphene had been put under the spotlight and enjoyed its several advantages of its fundamental properties. Despite the short history of graphene research, it has already revealed a series of new physics and potential applications, and no longer requires any further proof of its importance in quantum physics, condensed matter, and electronic devices.[1-4] However, the performance of graphene based electronic device is barely satisfactory. The zero bandgap of single layer graphene limits its possible applications in electronics even the carrier mobility of graphene can be reached up to $10^6$ cm$^2$/Vs.[5-9] Alternatively, transition metal dichalcogenide (TMD) is another material family with layered structure.[10-13] TMD family materials are composed of strong X-M-X interlayer covalent bonds, where X indicates the transition metal Mo or W; and X represents Se, S, or Te.[14-15] Similar to graphene, the bonding between different layers is the van der Waals force, showing the weak interlayer interactions where the isolation of single layer can be achieved by standard scotch tape method.[16] Molybdenum disulfide, MoS$_2$, one of the most studied TMD family materials, has been regarded as a promising candidate for field-effect transistors with relatively high on/off ratio and reasonable electron mobility.[16-28] With recent observation of the indirect into direct bandgap transition, MoS$_2$ based optoelectronic devices has attracted newest interest in optical society.[29-38] In addition, due to the atomically-thin, flexible, and bio-compatible nature of MoS$_2$, a completely new generation of electronic sensor devices can be envisioned[39-45]. Moreover, all those devices are based on individual MOSFETs, giving more demands on single transistor performance. In order to realize high-performance MoS$_2$ MOSFET and others, fundamental device physics of MoS$_2$ transistor is introduced first by clearly understanding of the switching mechanism of a Schottky barrier transistor. Different approaches to reduce the contact resistance on 2D materials are reviewed and explored in the latter part of this review. Record low contact resistance and high drain current are achieved on both $MoS_2$ and $WS_2$ after effective molecule chemical doping technique.

The nature of $MoS_2$ transistor is a Schottky barrier transistor, where the on/off states are switched by the tuning of the Schottky barriers at contacts.[46] As shown in Figure 1, we have two metal contacts that serve as source and drain for a single $MoS_2$ FET, named as source barrier and drain barrier. The effective barrier heights for source and drain barriers are primarily controlled by gate and drain biases. The carriers path for n-type $MoS_2$ transistor has been defined from the source to drain, that the electrons would encounter the source barrier first, where the carriers would undergo a thermal-assisted tunneling process from the source metal Fermi-level to the channel. On the other hand, the electrons in the channel would go from conduction band back to drain metals. Notably, gate bias has an opposite impact on these two barriers. As the high gate bias applied, the effective barrier height for source barrier, $\Phi_s$, is reduced due to a sharper triangle at the source end, where the effective Schottky barrier height has been shrank. However, lowering the conduction band at the large gate bias also enhances the $\Phi_d$, the effective Schottky barrier height at the drain end. With further increase of the drain bias, the barrier at the source end remains constant. However, the drain barrier starts to vanish with large magnitude of drain bias, facilitating electron carriers movement from the source to the drain.

Field-effect transistor built on ultra-thin few-layer MoS$_2$ is effectively the ultra-thin body FET, which has an optimal structure to immune the short channel effects.[47-49] Moreover, the heavier effective mass of the MoS$_2$ allows its transistors to have increased drive current, and enhanced transconductance when benchmarked against the ultrathin body Si transistors at their scaling limit.[50] Previous studies of MoS$_2$ transistors channel length scaling has aggressively pushed the channel length down to 50 nm,[51] where the device has demonstrated an inspiring characteristic in driving current, as shown in Figure 2(a). However, drain current saturation at the short channel regions had also been observed, which is directly attributed to the large contact resistance. The substantial contact resistance does not scale with channel length but remains almost same in the devices. As the channel length scales down to short channel regime, the channel resistance becomes comparable to the sum of two contact resistances. With further decrement of channel length, however, would not result in a significant improvement of drain current, where the drain voltage has been mainly applied on the two contacts.[19,51] Maximum drain current varies with different channel lengths of MoS$_2$ FETs had been reported in Figure 2(b). In the long channel regime, MoS$_2$ transistors have followed the classical square-law model that the drain current is inversely proportional to the channel length, $I_d \sim 1/L_{ch}$. With continuous channel length scaling down, the driving current starts to have a saturation at ~90 mA/mm at L = 100nm, which is due to the dominant contact resistance at short channel regime, indicating the sum of two contact resistances is comparable or even larger than the channel resistance.

Although, MoS$_2$ has attracted great interest for transistor applications because its large bandgap allows for field effect devices with low off-current, however, one key bottleneck in MoS$_2$ based device is the realization of the low-resistivity Ohmic contact. The on-state performance of the short channel MoS$_2$ FETs is mainly limited by its large contact resistance formed by Schottky barriers at the MoS$_2$/metal interfaces.[52-58] In this paper, we present a review of several approaches to reduce the contact resistance of MoS$_2$ and other 2D TMDs field-effect transistors, and improve device performance. By systemically analyzing the contact strategy among the 2D semiconducting materials, a roadmap for future high-performance TMDs FETs with low contact resistance is nearly approached.

The first approach which is widely studied now is to choose the low workfunction contact metals. Once the metal workfunction is close to the conduction band edge of the 2D materials, low resistivity contacts are expected. However, a number of recent articles have applied large work function metals, such as Ni or Au, as the contact metals on MoS$_2$ field-effect transistors, and yet reported decent n-type contact formation and drain current.[16,59,60] A detailed examination with different workfunction contact metals on MoS$_2$ field-effect transistors has been revealed to realize the heavily pinning phenomenon at the MoS$_2$/metal interface.[51,61-63] With different workfunction contact metals, n-type $MoS_2$ transistors had demonstrated a similar value of drain current and contact resistance, indicating a heavily pinning effect near the conduction band edge of $MoS_2$.[17-19,61-63] The band diagram showing different workfunction alignment of the contact metals to $MoS_2$ has been presented in Figure 3(a). Even though the contact metals with high workfunction are naturally close to the valence band of $MoS_2$, n-type transistor behaviors had all been realized in Ni, Au, and Pd contact metals, indicating a coincided Fermi-level pinning position at the semiconductor interface due to sulfur vacancies.[62] Different contact metal Fermi-level pinning positions had been shown in Figure 3(b), where both low and high workfunction metals are aligned near the conduction band of $MoS_2$, resulting in a monotonously n-type electrical characteristic. Experimental study of contacts to $MoS_2$ using low work function metal scandium (Sc) has been conducted to form an improved contact with $MoS_2$ film, which helps the electron carrier injection and to lower contact resistances for n-type $MoS_2$ transistors.[63]

In addition, recent studies in $MoS_2$ contact had revealed that the device characteristics can be changed to p-type Schottky barrier FETs using extreme high workfunction contact materials. Substoichiometric molybdenum trioxide ($MoO_{3-x}$), a high workfunction material aligned deeply into the valence band of $MoS_2$, has been demonstrated as a promising p-contact for $MoS_2$ transistor with a moderate drain current.[64] $MoO_{3-x}/MoS_2$ interface has obvious different pinning properties compared to metal/$MoS_2$ interfaces.[64]

Since both $MoS_2$ and graphene have the hexagonal structure with no dangling bands on surface, 2D to 2D interface may not have the traditional Fermi-level pinning issue. If the Fermi-level of graphene can move beyond the conduction band edge of $MoS_2$, graphene contact could enhance the electron injection and reduce the contact resistance to $MoS_2$. The second approach to improve the contact is to combine graphene and $MoS_2$ so as to create heterostructures which becomes an inspiring process to highlight the positive properties of each individual material.[65-75] N-type few-layer $MoS_2$ field-effect transistors with graphene/Ti as the hetero-contacts had been fabricated recently, where the chemical vapor deposition (CVD) grown monolayer graphene had been successfully transferred to the exfoliated $MoS_2$ contact area, and the contact resistance of hetero-contacts has been significantly reduced compared to metal-direct contact.[76,77] A maximum of 161.2 mA/mm drain current at 1 μm gate length with an on-off current ratio of $10^7$ had been achieved in few-layer $MoS_2$ field-effect transistors with graphene/Ti hetero-contacts. The enhanced electrical characteristic is confirmed in a nearly 2.1 times improvement in on-resistance and a 3.3 times improvement in contact resistance with hetero-contacts compared to the $MoS_2$ FETs without inserting graphene layer, as shown in Figure 4(a) and (b). The contact resistance is extracted from the transfer length method (TLM), which is a well known classical method for measuring the sheet and contact resistance at Ohmic or low resistivity limit. The TLM structure allows the measurements of various length resistors, and the resistance values can be plotted against the different resistor lengths. Sheet resistance and twice the contact resistance can be extracted from the slope and y-intercept values of the fitting curve, respectively.[78] The pronounced reduction in contact resistance from 12.1±1.2 Ω·mm to 3.7±0.3 Ω·mm with hetero-contacts structure is attributed to the gate-induced electron injection from graphene layer into $MoS_2$. In order to have fair comparison, contact resistances $R_c$ mentioned in this review had all been normalized by multiplying the channel width, where the unit of contact resistance is in Ω·mm. At the hetero-contacts structure, the positive back-gate bias not only electro-statically dopes $MoS_2$, but also could move the Fermi-level in Ti doped n-type graphene[79-81] further up beyond the Ti/$MoS_2$ pinning level, thus enhance the electron injection from metal into the conduction band of $MoS_2$. With advances in total carrier density summed over graphene and $MoS_2$ hetero-contacts exceeds the single $MoS_2$/Ti contact, the hetero-contact has demonstrated a lower contact resistance compared to traditional Ti-direct $MoS_2$ transistor. The similar improvements for the $MoS_2$ transistors with graphene/Pd hetero-contacts had been achieved as well,[77] where the output curves for both hetero-contacts and Pd-direct contact are shown in Figure 4(c). In particular, the 1 μm channel length graphene/Pd hetero-contacts device reached a drive current of 71.7 mA/mm, which is nearly 2.2 times higher than the direct Pd contact device with the same channel length. The hetero-contacts device lowers the linear regime FET on-resistance from 127.7 Ω·mm to 54.1 Ω·mm. Contact resistance for hetero-contacts and Pd-direct contact, extracted from TLM structure, have also been presented in Figure 4(d), where the contact resistance of graphene/Pd hetero-contacts dropped to 5.2 Ω·mm, compared with 23.5 Ω·mm for Pd-direct contact devices. In graphene/Pd hetero-contacts, the Fermi-level in Pd doped p-type graphene has been pushed up across Dirac point by electro-statically doping from back gate bias,[82] where the electrons can be injected from graphene into MoS$_2$ conduction band at high bias regions, resulting in a significant contact resistance reduction. Fermi-level of graphene under metal Pd is not pinned so as on MoS$_2$. Unpinned 2D interfaces provide unprecedented opportunities to explore new device structures. Similarly, high-performance devices with hereto-contacts of heavily doped graphene had also been achieved in MoS$_2$ based logic circuit[83] and semiconducting TMD FETs with ion gel[84] very recently.

The third approach is to heavily dope the source/drain regions of 2D materials. Heavily doped channel would significantly reduce Schottky barrier width thus reduce the contact resistance. Engineering electronic performance via doping is still in its infancy for MoS$_2$. Due to its nature of ultra-thin body structure, MoS$_2$ may not be doped as Si and III-V semiconductors by heavy ion implantation method; however, the ultra thin body nature allows the exploration of novel approaches, such as solid doping,[85] gas doping,[86] molecular doping,[87,88] and chemical doping.[89,90] One of organic chemicals, polyethyleneimine (PEI), has been proved to be an effective doping molecule in MoS$_2$ field-effect transistors.[87] The amine-rich aliphatic polymer, PEI is a widely used n-type surface dopant, for doping low dimensional nano-materials devices due to its strong electron-donating ability.[91-94] As shown in Figure 5(a) and (b), both channel resistance and contact resistance of few-layer MoS$_2$ field-effect transistors have been lowered after the application of PEI molecules. Channel resistance measured after PEI doping is 7.7±1.8 kΩ/□, decreased from 20.0±3.3 kΩ/□ before doping. This nearly 2.6 times reduction in channel resistance is attributed to the electrons transfer from the PEI molecules to the MoS$_2$ flake, where the PEI acts like a "charge donor". Contact resistance measured after PEI treatment had dropped to 4.6±1.1 Ω·mm, compared with 5.1±1.7 Ω·mm without PEI application. Approximately 20% lowering of contact resistance can be attributed to the reduction of Ti-MoS$_2$ Schottky barrier width for electron injection. As depicted in Figure 5(c), MoS$_2$ FETs on-current increases from 10.25 mA/mm to 17.61 mA/mm, where the 70% enhancement in on-current is achieved. Also, extrinsic field-effect mobility before and after PEI doping are calculated to be 20.4 and 32.7 cm$^2$/Vs for a typical 3 μm channel length device. Figure 5(d) shows the output characteristics of MoS$_2$ transistors before and after PEI applied. The improvement of current illustrates that PEI doping successfully improves the device performance of MoS$_2$ n-type transistors. However, PEI doping on MoS$_2$ is not very stable because of the charge transfer nature, which degrades with time.[87]

An important breakthrough on 2D materials contact research has been achieved very recently by substitute doping of Cl using 1,2 dichloroethane (DCE). A record low contact resistance of 0.5 Ω·mm and record high drain current of 460 mA/mm had been reported for the first time in MoS$_2$ device field.[89] The strong n-type doping phenomenon in DCE doped MoS$_2$ could be ascribed to the donation of extra electron when substitution of S$^{2-}$ by Cl$^-$ takes place, particularly at the sites of sulfur vacancies in the MoS$_2$ thin film. As shown in Figure 6(a), contact resistance extracted from TLM is significantly reduced from 5.4 Ω·mm to 0.5 Ω·mm after the Cl doping. The improvement in contact resistance can be attributed to the doping induced thinning of tunneling barrier width. Also, the transfer length (L$_T$) of metal-MoS$_2$ junctions is determined to be 60 nm and 590 nm for the contacts with and without the Cl doping, respectively. Shown in Figure 6(b), compared with the control sample without the Cl doping, the contact resistivity $\rho_c$, where $\rho_c = R_c L_T$ is in the unit of $\Omega \cdot cm^2$, is reduced from $3 \times 10^{-5}$ $\Omega \cdot cm^2$ to $3 \times 10^{-7}$ $\Omega \cdot cm^2$, which is comparable to traditional Si and III-V semiconductor contact interface. Output characteristics of a 100 nm channel length $MoS_2$ FET with and without the Cl doping is depicted in Figure 6(c). The reduced contact resistance helps to enhance the drain current from ~ 110 mA/mm up to 460 mA/mm, which is twice of the best reported value so far on $MoS_2$ FETs at the same channel length. The transfer behaviors of both doped and undoped $MoS_2$ FETs are shown in Figure 6(d). A $6.3 \times 10^5$ on/off ratio with 50-60 $cm^2$/Vs field-effect mobility has been approached in Cl doped $MoS_2$ transistors.

The demonstrated chloride molecular doping technique can also be expanded to other TMDs 2D materials.[95] One good example is $WS_2$ where the doping effect can be even more highlighted than $MoS_2$ case. The mechanism of the reduction of the $R_c$ on Cl-doped TMDs is very clear. The energy band diagrams of the Ni-$WS_2$ and Ni-$MoS_2$ contacts with and without the Cl doping are shown in Figure 7(c). The Fermi-level at the metal-$WS_2$ interface is pinned near the charge neutrality level (CNL), resulting a significantly large Schottky barrier. The height of the Schottky is large enough to rectify the electrons' ejection from the metal to the semiconductor at low $V_{ds}$, as shown in Figure 7(c). Moreover, this barrier height can't be efficiently modified by varying the workfunction of contact metals due to the complicated metal-to-TMD interface as described above. The difference of the $R_c$ between $WS_2$ and $MoS_2$ is due to the different alignment of the CNL in the two materials. Compared with $MoS_2$, the CNL in $WS_2$ is more close to the middle of the bandgap, resulting in a larger Schottky barrier. Without doping, it would be much harder for the electrons to inject from the metal to the semiconductor in $WS_2$ because the thermionic current exponentially decreases with the increasing of barrier height. However, when the tunneling current starts to dominate the current through the M-S junction, the electron injection through the barrier becomes much easier. The effective electron density (induced by chemical doping and electrostatic doping) at $V_{bg}$ of 50 V is as high as $2.3 \times 10^{13}$ cm$^{-2}$ and $2.9 \times 10^{13}$ cm$^{-2}$ for $WS_2$ and $MoS_2$, respectively.[95] As a result, both of the $R_c$ in the $WS_2$ and $MoS_2$ decrease significantly after doping. However, it is interesting to note that most of the electron density in $WS_2$ is attributed to the back gate bias rather the chemical doping because the electron density of $WS_2$ is determined to be only $6.0 \times 10^{11}$ cm$^{-2}$ at zero back gate bias.[95] In another word, the Fermi-level (electron density) at the interface can be effectively modulated by the back gate bias. Effective modulation via field-effect can be ascribed to the passivation of sulfur vacancy by Cl, given that the sulfur vacancy is the cause of the Fermi-level pinning on $MoS_2$ and $WS_2$ at M-S interface.

The $R_c$ of $WS_2$ can be significantly reduced after the Cl doping. Known as an ambipolar semiconductor, the undoped $WS_2$ shows large Schottky barriers for both electrons and holes, resulting an extremely large $R_c$.[96] For such a larger Schottky barrier, it would be impractical to extract the $R_c$ by the TLM structure which is applicable to Ohmic or low resistivity contacts only. However, a simple estimation of the $R_c$ of the undoped $WS_2$ is on the order of $10^2$ Ω·mm since the total resistance of the 100 nm device is calculated to be $5\times10^2$ Ω·mm. After doping, an $R_c$ as low as 0.7 Ω·mm, 2-3 orders of magnitude reduction, can be extracted by linearly fitting the curve of total resistances. Figure 7(a) shows the TLM resistances of the Cl-doped $WS_2$ as a function of gap space at a back gate bias of 50 V. Since the low $R_c$ is achieved in $WS_2$ by Cl doping, high-performance $WS_2$ FET is expected. The output characteristics of the Cl-doped few-layer $WS_2$ FETs with 100 nm channel length are shown in Figure 7(b). The device exhibits promising device performance including a drain current of 380 mA/mm as well as good current saturation. Due to a small $R_c$, the linear region of the $I_{ds}$-$V_{ds}$ curves shows excellent linearity. The drain current starts to saturate at $V_{ds}$ of 1.0 V due to the electron velocity saturation. To the best of our knowledge, such a low $R_c$ and a large drain current have never been achieved on $WS_2$ or other TMDs whose CNL is located in the middle of the bandgap.

With advances of its ultra-thin body, decent mobility and sizable bandgap, $MoS_2$ has been regarded as a typical semiconducting 2D material for the next generation channel material for thin-film transistor technology. However, one of the major road blocks for high-performance $MoS_2$ transistors is the exhibiting Schottky barrier at the metal/semiconductor interface thus large contact resistance. In this review, $MoS_2$ device physics had been firstly introduced to understand the contact involved switching mechanism in $MoS_2$ FETs. More importantly, contact research strategies to reduce $R_c$ on $MoS_2$ and other 2D TMDs transistors had been elucidated to help realization of the high-performance 2D FETs with low $R_c$ for future electronics applications.

**Acknowledgement**

The work is supported by SEMATECH and SRC. The authors would like to particularly thank Kausik Majumdar from SEMATECH and Wilman Tsai from Intel Corporation for the valuable discussions and technical assistance.

**Figures**

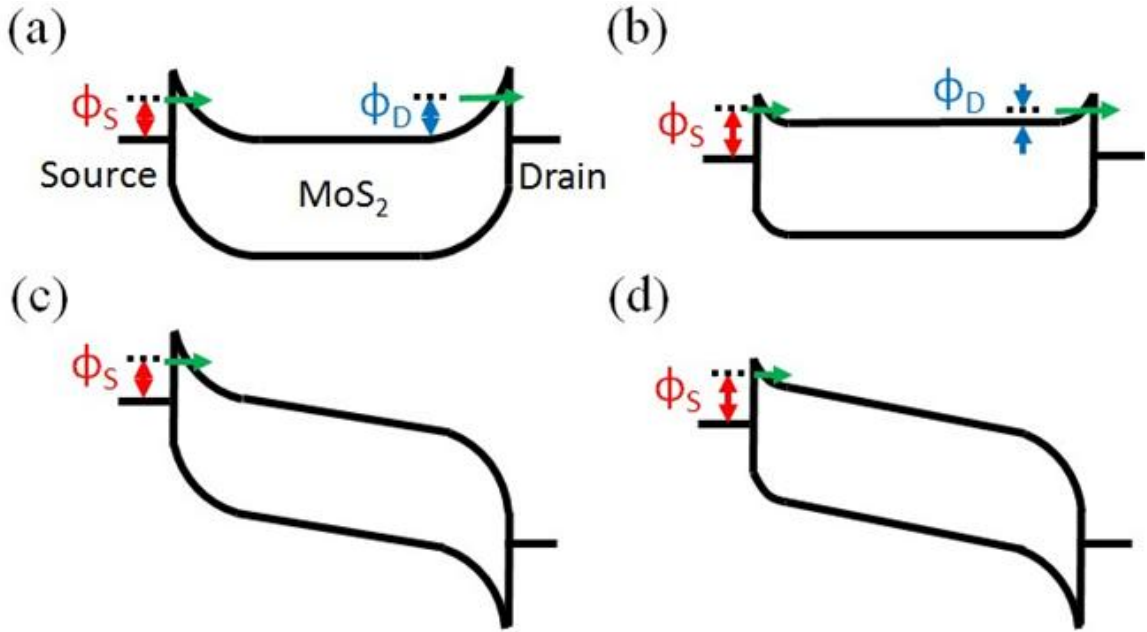

Figure 1. (a) Band diagram with large back gate bias and zero drain bias. (b) Band diagram with small back gate bias and zero drain bias. (c) Band diagram with large back gate bias and large drain bias. (d) Band diagram with small back gate bias and large drain bias. Reprinted with permission from H. Liu, M. Si, Y. Deng, A. T. Neal, Y. Du, S. Najmaei, P. M. Ajayan, J. Lou, and P. D. Ye, ACS Nano 8, 1031 (2014). Copyright 2014, American Chemical Society.

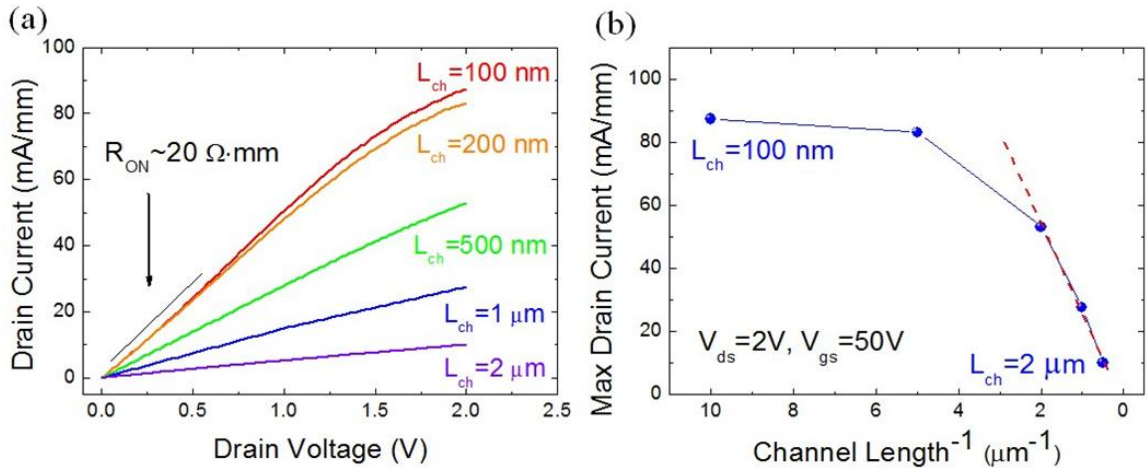

Figure 2. (a) Output curves of different channel lengths transistors at the same back gate voltage condition. (b) Magnitude of on-state drain current varies with different channel length, drain current saturation has been observed in the short channel regime. Reprinted with permission from H. Liu, A. T. Neal, and P. D. Ye, ACS Nano 6, 8563 (2012). Copyright 2012, American Chemical Society.

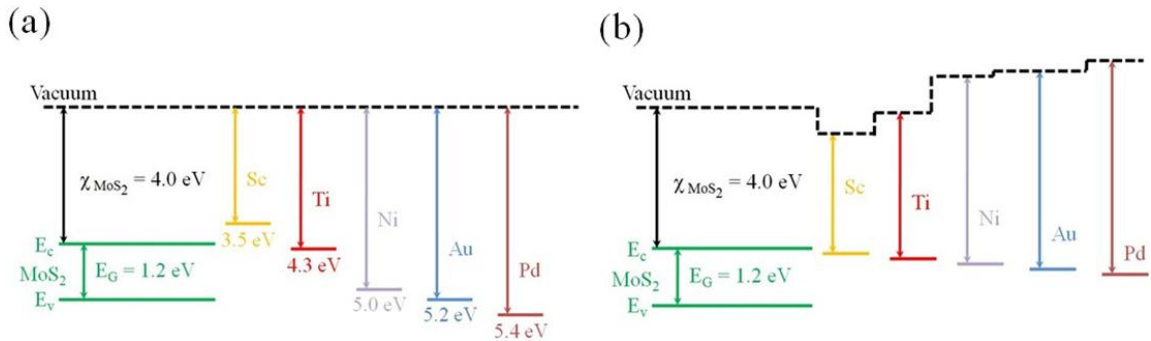

Figure 3. (a) Band diagram showing the different workfunction alignment of the contact metals to $MoS_2$. (b) Band diagram showing the different workfunction pinning position of the contact metals to $MoS_2$.

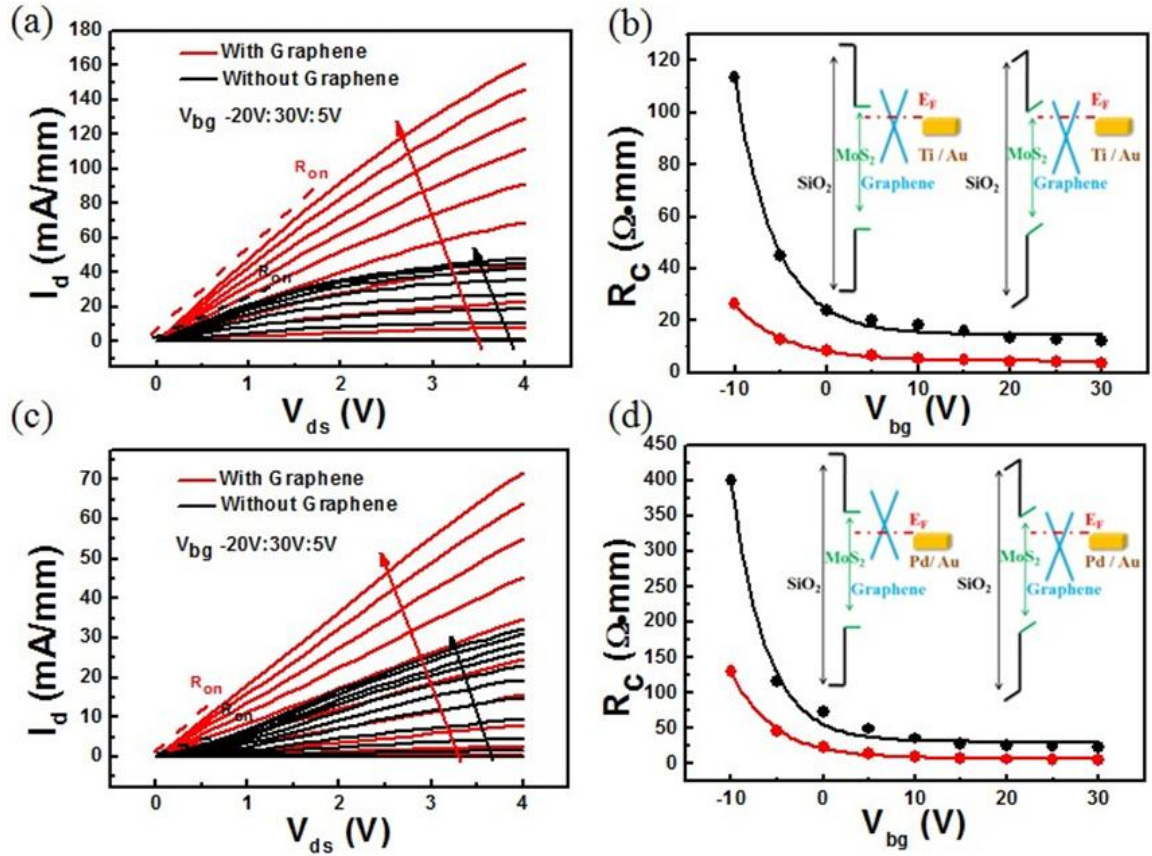

Figure 4. (a) Output characteristics for MoS$_2$ Field-effect transistor with hetero-contacts and Ti-direct contact. (b) Contact resistance versus back-gate voltage for hetero-contacts and Ti-direct contact. (Inset: band alignment of Ti/graphene/MoS$_2$ contact under zero and positive gate bias). (c) Output characteristics for MoS$_2$ Field-effect transistor with hetero-contacts and Pd-direct contact. (d) Contact resistance versus back-gate voltage for hetero-contacts and Pd-direct contact. (Inset: band alignment of Pd/graphene/MoS$_2$ contact under zero and positive gate bias).

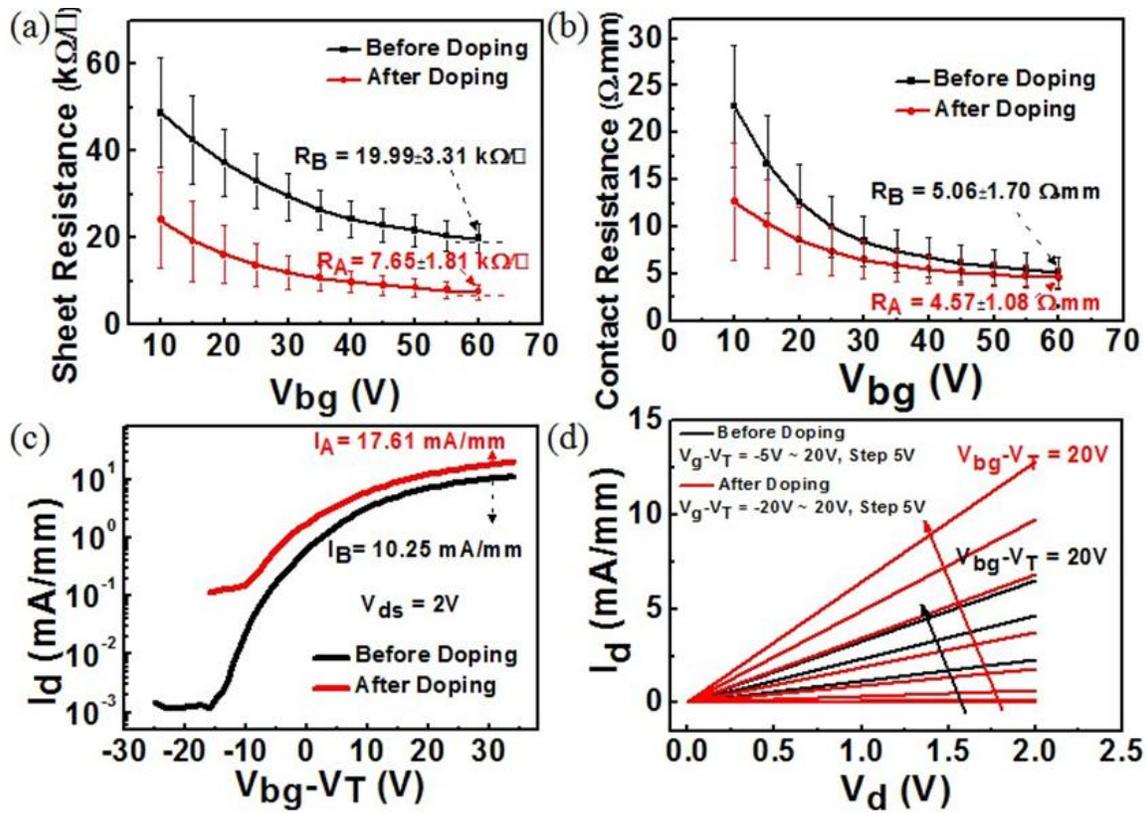

Figure 5. (a) Channel resistance before and after PEI doping varies with different back gate bias. (b) Contact resistance before and after PEI doping varies with different back gate bias. (c) Transfer characteristics of a channel length L = 3μm transistor before and after PEI doping. (d) Output characteristics of the same transistor before and after PEI doping. Reprinted with permission from Y. Du, H. Liu, A. T. Neal, M. Si, and P. D. Ye, IEEE Electron Device Lett. 34, 1328 (2013). Copyright 2013, Institute of Electrical and Electronics Engineers.

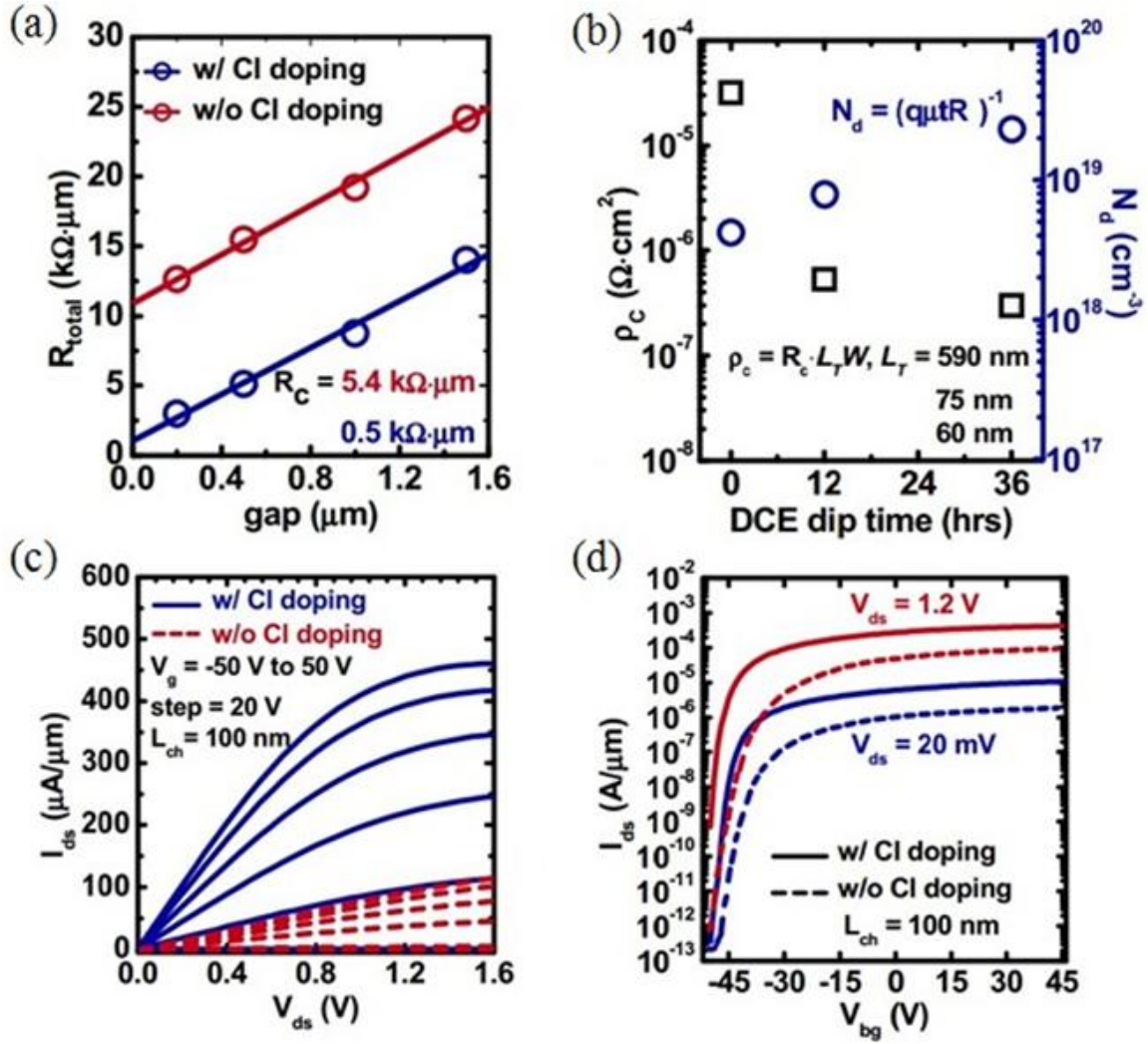

Figure 6. (a) TLM structure with total resistance varies with different channel length. Contact resistance extracted from the intercept shows a significant reduction from 5.4 Ω·mm to 0.5 Ω·mm. (b) Contact resistivity and carrier concentration for doped $MoS_2$ flake with different DCE dip time. (c) Comparison in output characteristics of a 100 nm channel length $MoS_2$ FET under Cl doping. A record high drain current of 460 mA/mm is obtained as Cl doping applied. (d) Comparison in transfer characteristic curves of a 100 nm channel length $MoS_2$ FET under Cl doping. Reprinted with permission from L. Yang, K. Majumdar, Y. Du, H. Liu, H. Wu, M. Hatzistergos, P. Hung, R. Tieckelmann, W. Tsai,

C. Hobbs, and P. D. Ye, 2014 Symposium on VLSI Technology, 238 (2014). Copyright 2014, Institute of Electrical and Electronics Engineers.

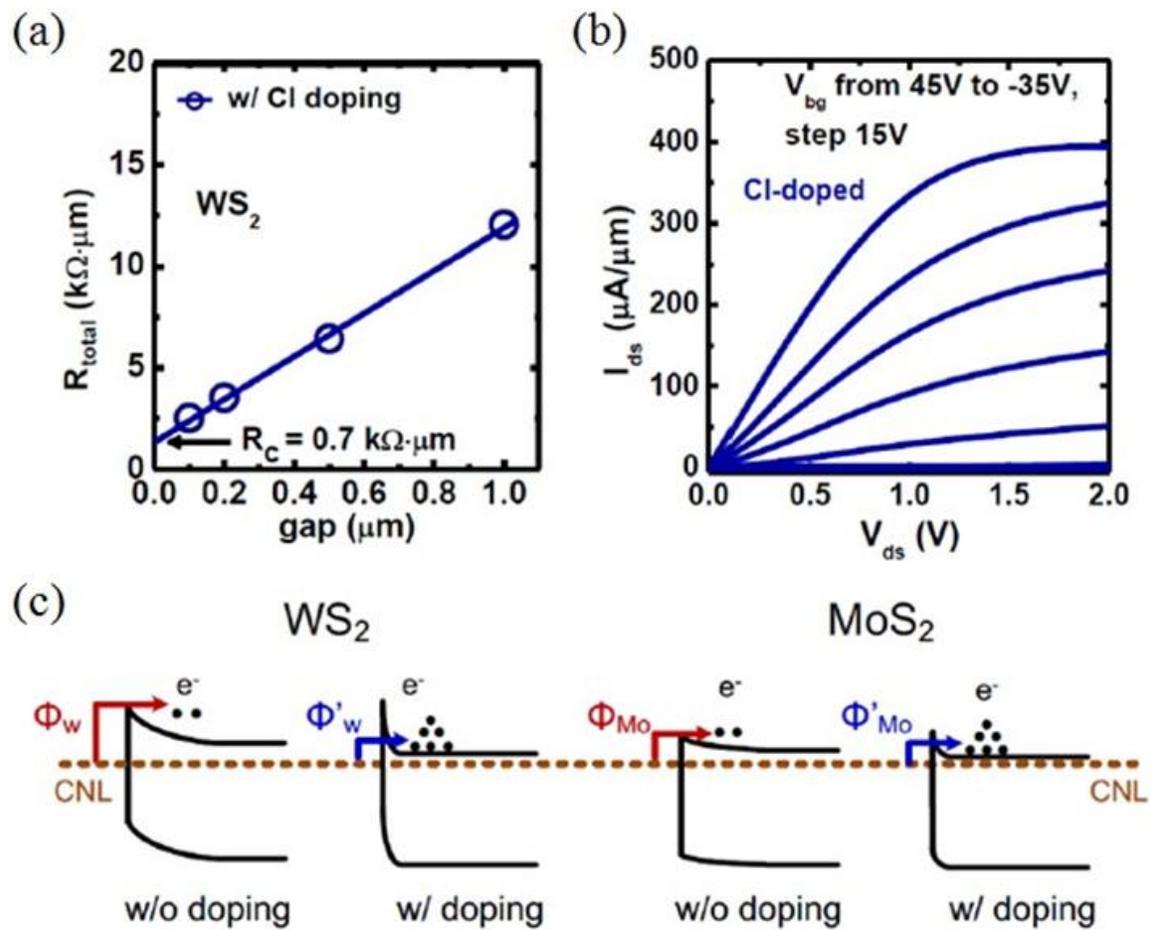

Figure 7. (a) TLM resistances of Cl-doped $WS_2$. The $R_c$ is extracted to be 0.7 Ω·mm. The $L_T$ is extracted to be 132 nm and the corresponding $\rho_c$ is about $9.2 \times 10^{-6}$ Ω·cm$^2$. The doping density is about $6.0 \times 10^{11}$ cm$^{-2}$. (b) Output characteristics of the Cl-doped few-layer WS$_2$ FETs at 100 nm channel length. The maximum drain current is enhanced to 380 mA/mm after Cl doping. Excellent current saturation is also observed. (c) Schematic band diagram of metal-TMD contacts with and without chloride doping. Before DCE treatment, the Fermi-level is pinned close to the CNL, resulting in a large Schottky barrier. After DCE treatment, WS$_2$ and MoS$_2$ are heavily doped and the Fermi-level in 2D materials can be efficiently moved after the passivation of S vacancy by Cl dopants.